\documentclass[useAMS,usenatbib]{mn2e}

\usepackage{color}
\usepackage{graphicx}
\usepackage{amssymb}
\usepackage{amsmath}
\usepackage{hyperref}
\usepackage{threeparttable}
\usepackage{tabularx}

\title[H{\sc i} study of the environment around the galaxy ESO 243-49]{H{\sc i} study of the environment around ESO 243-49, the host galaxy of an intermediate mass black hole}
\author[A. Musaeva et al.]{\parbox{\textwidth}{A. Musaeva$^{1,2,3}$\thanks{E-mail: a.musaeva@physics.usyd.edu.au}, B. S. Koribalski$^{3}$, S. A. Farrell$^{1}$, E. M. Sadler$^{1}$, M. Servillat$^{4,5}$, R. Jurek$^{3}$, E. Lenc$^{1,2}$, R. L. C. Starling$^{6}$, N. A. Webb$^{7,8}$, O. Godet$^{7,8}$, F. Combes$^{9}$ and D. Barret$^{7,8}$}
\vspace{0.4cm}\\
\parbox{\textwidth}{$^{1}$Sydney Institute for Astronomy, School of Physics, University of Sydney, Redfern, NSW, 2016, Australia\\
$^{2}$ARC Centre of Excellence for All-sky Astrophysics (CAASTRO), Redfern, NSW, 2016, Australia\\
$^{3}$Australia Telescope National Facility, CSIRO Astronomy and Space Science, PO Box 76, Epping, NSW 1710, Australia\\
$^{4}$Harvard-Smithsonian Center for Astrophysics, 60 Garden Street, Cambridge, MA 02138, USA\\
$^{5}$Laboratoire AIM (CEA/DSM/IRFU/SAp, CNRS, Universit\'{e} Paris Diderot), CEA Saclay, Bat. 709, 91191 Gif-sur-Yvette, France\\
$^{6}$Department of Physics and Astronomy, University of Leicester, University Road, Leicester LE1 7RH, UK\\
$^{7}$Institut de Recherche en Astrophysique and Plan\'{e}tologie (IRAP), Universit\'{e} de Toulouse, UPS, 9 Avenue du colonel Roche, F-31028 Toulouse Cedex 4, France\\
$^{8}$CNRS, UMR5277, F-31028 Toulouse, France\\
$^{9}$LERMA, Observatoire de Paris, UMR 8112 61, Av. de l'Observatoire, F-75014 Paris}
}
\begin{document}

\date{Accepted 2014 December 2. Received 2014 November 17; in original form 2014 October 15}

\pagerange{\pageref{firstpage}--\pageref{lastpage}} \pubyear{2014}

\maketitle

\label{firstpage}

\begin{abstract}
The lenticular galaxy ESO 243-49 hosts the ultraluminous X-ray source HLX-1, the best candidate intermediate mass black hole (IMBH) currently known. The environments of IMBHs remain unknown, however the proposed candidates include the nuclei of dwarf galaxies or globular clusters. Evidence at optical wavelengths points at HLX-1 being the remnant of an accreted dwarf galaxy. Here we report the Australia Telescope Compact Array radio observations of H{\sc i} emission in and around ESO 243-49 searching for signatures of a recent merger event. No H{\sc i} line emission is detected in ESO 243-49 with a 5$\sigma$ upper limit  on the H{\sc i} gas mass of a few $10^8 M_{\odot}$. A likely reason for this non-detection is the cluster environment depleting ESO 243-49's H{\sc i} gas reservoir. The upper limit is consistent with an interpretation of HLX-1 as a dwarf satellite of ESO 243-49, however more sensitive observations are required for a detection.
\par We detect $\sim$5$\times 10^8 M_{\odot}$ of H{\sc i} gas in the peculiar spiral galaxy AM 0108-462, located at a projected distance of $\sim$170 kpc from ESO 243-49. This amount of H{\sc i} gas is $\sim$10 times less than in spiral galaxies with similar optical and near-infrared properties in the field, strengthening the conclusion that the cluster environment indeed depletes the H{\sc i} gas reservoir of these two galaxies. Here we also report observations of AM 0108-462 in several optical and near-infrared bands using the Magellan 6.5 m telescopes, and archival X-ray and ultraviolet observations with \textit{XMM-Newton} and \textit{Swift}. These data combined with the H{\sc i} line data suggest it is likely that AM 0108-462 is experiencing a merger event.

\end{abstract}


\section{Introduction}

Most black holes we observe today fall into two categories: supermassive black holes (SMBHs) with masses in the range $10^6$-$10^9 M_{\odot}$ residing in the centres of many large galaxies \citep[e.g.,][]{Kormendy1995}, and stellar mass black holes with masses $<$100 $M_{\odot}$, formed through the collapse of massive stars \citep[e.g.,][]{Belczynski2010}. Despite very strong observational evidence for the existence of SMBHs, we don't understand how to form them, although there are several possible formation mechanisms \citep{Rees1978}. In one theory the SMBHs are thought to be formed from lighter mass black hole seeds, which are either remnants of the first generation of stars $\left(\sim10^2M_{\odot} \right)$ or collapsed gas clouds $\left(\gtrsim10^3M_{\odot}\right)$ \citep{Volonteri2010}. The seeds must then grow very rapidly since we observe luminous quasars powered by SMBHs in the first Gyr after the Big Bang \citep[e.g.,][]{Barth2003, Willott2005}. Another theory proposes that several IMBHs with masses between $10^2$ and $10^5 M_{\odot}$ merge in the center of a dense cluster of stars \citep{Ebisuzaki2001}. Following such merger events the rapid growth to the SMBH is possible through a super-Eddington accretion phase \citep[e.g.,][]{Kawaguchi2004}. Either way, IMBHs are predicted to be an integral phase of the SMBH formation and are thus important for understanding the formation and evolution of SMBHs and their host galaxies, however, convincing evidence for the existence of IMBHs has been lacking.

\par The best IMBH candidates have been proposed to be ultraluminous X-ray sources (ULXs) - extragalactic X-ray sources located away from the nucleus of the host galaxy with bolometric luminosities exceeding the Eddington limit of stellar mass black holes \citep[i.e. $>10^{39}$ erg s$^{-1}$ for a 10 $M_{\odot}$ black hole; e.g.,][]{Feng2011}. Assuming isotropic emission below the Eddington limit, these luminosities imply the presence of an accreting black hole with a mass of $\sim$$10^2$-$10^5 M_{\odot}$. However, most ULXs can be explained as stellar mass black holes either undergoing hyper-accretion or beamed emission, causing them to reach apparent luminosities of up to 10 times the Eddington limit. Taking an upper limit of 100 $M_{\odot}$ for stellar mass black holes therefore means that ULXs with luminosities above $\sim$$10^{41}$ erg s$^{-1}$ are difficult to explain without a more massive black hole.

\par Currently the best IMBH candidate is HLX-1, located in the edge-on \textit{S0a} galaxy ESO 243-49, $\sim$3.3 kpc from the nucleus and $\sim$1 kpc out of the plane \citep{Farrell2009}. A spectroscopic redshift measurement of H$\alpha$ emission at the position of HLX-1 is consistent with that of ESO 243-49, placing it at a distance of $\sim$95 Mpc \citep{Wiersema2010}. The maximum 0.2-10 keV unabsorbed X-ray luminosity (assuming isotropic emission) of $1.3\times10^{42}$ erg s$^{-1}$ is an order of magnitude larger than the previously known brightest ULX \citep[e.g.,][]{Gao2003}, $\sim$400 times the Eddington limit of a 20 $M_{\odot}$ black hole. The best constraints on the black hole mass place it between 9,000 and 90,000 $M_{\odot}$ \citep{Webb2012}. HLX-1's extreme luminosity rules out the possibility of a stellar mass black hole, therefore making it the best IMBH candidate currently known.

\par The nature of the environment around HLX-1 remains an open question. Farrell et al. \citeyearpar{Farrell2012} analysed \textit{Hubble Space Telescope} (\textit{HST}) and \textit{Swift} X-ray Telescope data of the optical counterpart of HLX-1 and showed (through fitting the spectral energy distribution from X-ray to near-infrared wavelengths) that the broadband spectrum was well described by a model comprised of an irradiated accretion disc plus a stellar population around HLX-1. They reported that the stellar population was likely to be a young $\left(\sim13\:\mathrm{Myr}\right)$ massive $\left(\sim4\times10^6M_{\odot}\right)$ star cluster. However, Farrell et al. \citeyearpar{Farrell2012} could not exclude the presence of a lower luminosity old population of stars ($\sim$10 Gyr) with a total stellar mass of up to $\sim10^6M_{\odot}$.

\par The optical counterpart of HLX-1 was observed with the \textit{Very Large Telescope} (\textit{VLT}) $\sim$2 months after the aforementioned \textit{HST} observations, and the optical flux was reported to drop by a factor of $\sim$2 \citep{Soria2012}. Soria et al. \citeyearpar{Soria2012} attributed most of the optical and ultraviolet emission to an irradiated accretion disc, concluding instead that if there was some contribution from a star cluster, it would either be a massive $\left(\lesssim2\times10^6M_{\odot}\right)$ old $\left(\sim10\:\mathrm{Gyr}\right)$ population or lower mass $\left(\sim10^4M_{\odot}\right)$ young $\left(\lesssim10\:\mathrm{Myr}\right)$ population of stars. Combining the data from the \textit{Swift} X-ray Telescope with the \textit{HST} and \textit{VLT} data, Farrell et al. \citeyearpar{Farrell2014} found that a model with an irradiated accretion disc with stellar component with age of $\sim$20 Myr and mass of $\sim10^5M_{\odot}$ produced the best fitting solution. However, the data did not exclude the presence - in addition to the population of young stars that dominate the stellar light - of a lower luminosity population of older $\left(\sim10\:\mathrm{Gyr}\right)$ stars with a total stellar mass up to $\sim10^6M_{\odot}$ \citep{Farrell2014}.

\par A young star cluster surrounding HLX-1 could be explained in the accreted dwarf galaxy scenario, in which a trail of star formation could be created by ram-pressure stripping of gas and stars from a dwarf galaxy which has recently interacted with ESO 243-49 \citep{Webb2010}. In this case HLX-1 could have been an IMBH which was once at the centre of the dwarf galaxy, but has now had most of the gas and stars stripped off via the gravitational interaction with ESO 243-49. The compact nucleus of the dwarf galaxy, during its passage through ESO 243-49, could also have collected gas from the main galaxy, and this could perhaps be fueling HLX-1 \citep{Soria2010}. The presence of dust lanes and the lack of nuclear activity from X-ray observations of ESO 243-49 might also suggest a gas-rich minor merger $<$200 Myr ago \citep{Farrell2012}. Mapelli et al. \citeyearpar{Mapelli2013} reported that the comparison between the \textit{HST} photometric data of ESO 243-49 and their simulations confirmed that a minor merger was a viable scenario to explain the properties of HLX-1 and of its optical counterpart.

\par These results continue to support the scenario whereby HLX-1 is the remnant of a dwarf galaxy that has been accreted and stripped of most of its mass through a merger event with ESO 243-49. In this framework, a higher total mass of the stellar cluster than that derived for the dominant population of young stars (thus implying that an older population of low luminosity stars may dominate the mass) is likely required in order to explain how HLX-1 managed to retain the gas necessary for the recent burst of star formation.

\par Neutral hydrogen (H{\sc i}) 21-cm spectral line observations are an excellent probe of the local and large-scale environment around HLX-1 via mapping of the redshifted H{\sc i} and radio continuum emission both in its host galaxy and in ESO 243-49's galaxy cluster. The detection of H{\sc i} gas can provide signatures of past galaxy interactions such as tidal tails, a distorted H{\sc i} disc, peculiar velocity field, and extra-planar gas. Although early type galaxies in clusters are typically devoid of H{\sc i} gas, H{\sc i} emission has been significantly detected around a number of \textit{S0}-type galaxies that have undergone recent mergers \citep[e.g.,][]{Emonts2008, Struve2010,Serra2012}. Detection of extra-planar H{\sc i} gas from such a merger event around the location of HLX-1 would strongly support the dwarf galaxy scenario.

\par ESO 243-49 is a member of Abell 2877 galaxy cluster having 89 optical cluster members. Abell 2877's central cluster dominant galaxy, IC 1633 is located at a projected distance of $\sim$0.3 Mpc from ESO 243-49. The cluster is `poor', having a richness class $R=0$. Abell 2877 is an X-ray source with a temperature of $\sim$3.5 keV, it shows no evidence for a cooling flow. Its virial radius is $\sim$1.8 $h^{-1}$ Mpc, a velocity dispersion of 900 km s$^{-1}$, and a virial mass of $\sim$5$\times 10^{14} h^{-1} M_{\odot}$\citep[][and references therein]{Hopkins2000}.

\par Here we report on the analysis of H{\sc i} spectral line and radio continuum observations of the field of ESO 243-49 in an attempt to test the scenario in which HLX-1 is the remnant of a dwarf galaxy that has been accreted and stripped of most of its mass through a merger event with ESO 243-49. Section \ref{secRadioObservations} describes the aforementioned observations and the data reduction procedures, while Section \ref{secResults} contains the results obtained for ESO 243-49 and AM 0108-462, a peculiar galaxy in the field of ESO 243-49, for which we analyse the additional data in several wavebands. Section \ref{secDiscussion} discusses the implications of these results, and Section \ref{secConclusion} summarises the conclusions we have drawn.

\section{Observations and Data Reduction}\label{secRadioObservations}

\par The radio continuum and H{\sc i} spectral line observations were obtained with the Australia Telescope Compact Array \citep[ATCA;][]{Frater1992}. The ATCA is a radio interferometer that consists of six dishes (each 22 m in diameter).  The five inner antennae of the array can be moved around on the track to give ten baselines of lengths between 31 m and 3 km. The CA06 antenna (positioned a further 3 km beyond the end of the track) is fixed providing five more baselines with a maximum length of 6 km.

\par We conducted 21-cm observations of the field of ESO 243-49 to search for H{\sc i} emission associated with this galaxy (all previous observations of the field have been continuum observations and primarily at higher frequencies \citep{Hopkins2003, Webb2012, Cseh2014}, hence are not suitable for the H{\sc i} spectral line study). We used three array configurations under the project C2703 (PI S. Farrell). Most of the observations were carried out during the day (due to the scheduling constraints), hence they were susceptible to solar radio frequency interference (RFI). The details of these observations are summarised in Table \ref{tblATCAObservations}. The EW352 configuration was chosen to provide maximum sensitivity to the total H{\sc i} gas and large-scale diffuse emission. The longer baselines of the 750A and 1.5B arrays allow for the study of H{\sc i} structures over small angular scales. The combination of all three arrays provides good coverage of the \textit{uv} plane while retaining sensitivity to diffuse emission and achieving sufficient spatial resolution to image clumpy H{\sc i} gas in and outside the disc of the galaxy.

\par We used the Compact Array Broadband Backend \citep[CABB;][]{Wilson2011} zoom modes, in the 64M-32k configuration. A 32-channel 2 GHz continuum band and a 2,048-channel 64 MHz zoom band were observed simultaneously in this configuration. The 64 MHz zoom band covers the range between 1,364 and 1,428 MHz, centred on 1,396 MHz. Each of the channels is 32 kHz wide, providing a spectral resolution of $\sim$7 km s$^{-1}$. The target field of ESO 243-49 was observed in 20 minute scans, interleaved with 3 minute scans of the nearby phase calibrator PKS 0131-522. The amplitude calibrator, PKS 1934-638, was observed for 10-15 minutes at the beginning of each of the three observations.

\begin{table}
\caption{Summary of ATCA Observations of ESO 243-49.}
	\begin{center}
		\begin{tabular}{lcl}
			\hline 
			\hline            
			Date&Time (hours)&Array Configuration\\
			\hline
 			April 18 2012&12&1.5B\\
			April 22 2012&9&EW352\\
			June 28 2012&12&750A\\
			\hline  
		\end{tabular}
	\end{center}
\label{tblATCAObservations}
\end{table}

\par For the data reduction and visualisation we used the Multichannel Image Reconstruction, Image Analysis and Display \citep[{\sc miriad};][]{Sault1995} and {\sc karma} software \citep[for signal and image processing and visualisation;][]{Gooch1996}. Flagging of RFI was done using the manual and interactive flagging packages {\sc uvflag} and {\sc blflag}. The ATCA primary calibrator PKS 1934-638 was used to calibrate the absolute flux density scale. In our data set PKS 1934-638's flux density was 14.92$\pm 0.05$ Jy. After flagging and calibration, we combined the visibilities from all three observations to produce radio continuum images and spectral line data cubes.

\par For continuum images we only used the 64 MHz zoom-band data since its division into 2,048 separate channels allowed for more selective RFI-flagging. The visibilities from all the antennae were used (including the CA06 6 km antenna) to achieve the highest spatial resolution. To minimise the thermal noise and maximise sensitivity we applied the natural weighting scheme by setting Briggs's visibility weighting robustness parameter to +2 \citep{Briggs1995}. The theoretical rms noise limit was 16$\mu$Jy beam$^{-1}$, and in our data set the rms noise level was 45$\mu$Jy beam$^{-1}$, partly due to solar and other RFI.

\par For the spectral line data cubes we subtracted the continuum from the line data in each individual data set by fitting a first-order polynomial to the line-free channels using {\sc uvlin}. The data from the CA06 6 km antenna were not used for creating H{\sc i} spectral line data cubes due to non-uniform coverage of the \textit{uv} plane. We produced spectral line cubes in the velocity range where the H{\sc i} emission was expected (6,700-7,200 km s$^{-1}$) using steps of 7 km s$^{-1}$. We applied the robust weighting scheme (by setting Briggs's visibility weighting robustness parameter to 0) because it improved the beam shape and minimised the thermal noise at the expense of only a small loss of sensitivity. The data set was then smoothed spatially by a factor of 2, applying Hanning smooth in velocity to achieve an rms noise level of 1.1 mJy beam$^{-1}$ (the theoretical rms noise limit was 0.9 mJy  beam$^{-1}$). Both continuum images and spectral line data cubes were primary beam corrected. To measure the fluxes of continuum sources we used the {\sc imfit} routine with a gaussian or point source component, and chose whichever was most appropriate for an individual source based on the values of the fitting parameters.

\par We calculated H{\sc i} masses by integrating the emission from all spectral channels according to the following equation \citep{Roberts1962}:
\begin{equation}
	M_{\mathrm{H{\textsc i}}}=2.36 \times 10^5 M_{\odot} \:d^2 F_{\mathrm{H{\textsc i}}},
	\label{HImass}
\end{equation}
where $d$ is the distance (in Mpc), and $F_{\mathrm{H{\textsc i}}}$ is the integrated flux (in Jy km s$^{-1}$). The `optical definition' and the barycentric standard-of-rest frame are used for all the velocity values reported in this study:
\begin{equation}
	v_{optical}=c\left(\frac{\nu_0}{\nu}-1\right),
	\label{OpticalVelocity}
\end{equation}
where $\nu$ is the observed frequency, $\nu_0$ is the rest frequency and $c$ is the speed of light.

\section{Results}\label{secResults}

\subsection{H{\sc i} and radio continuum data}

\par
\begin{figure*}
 \includegraphics[width=1\textwidth]{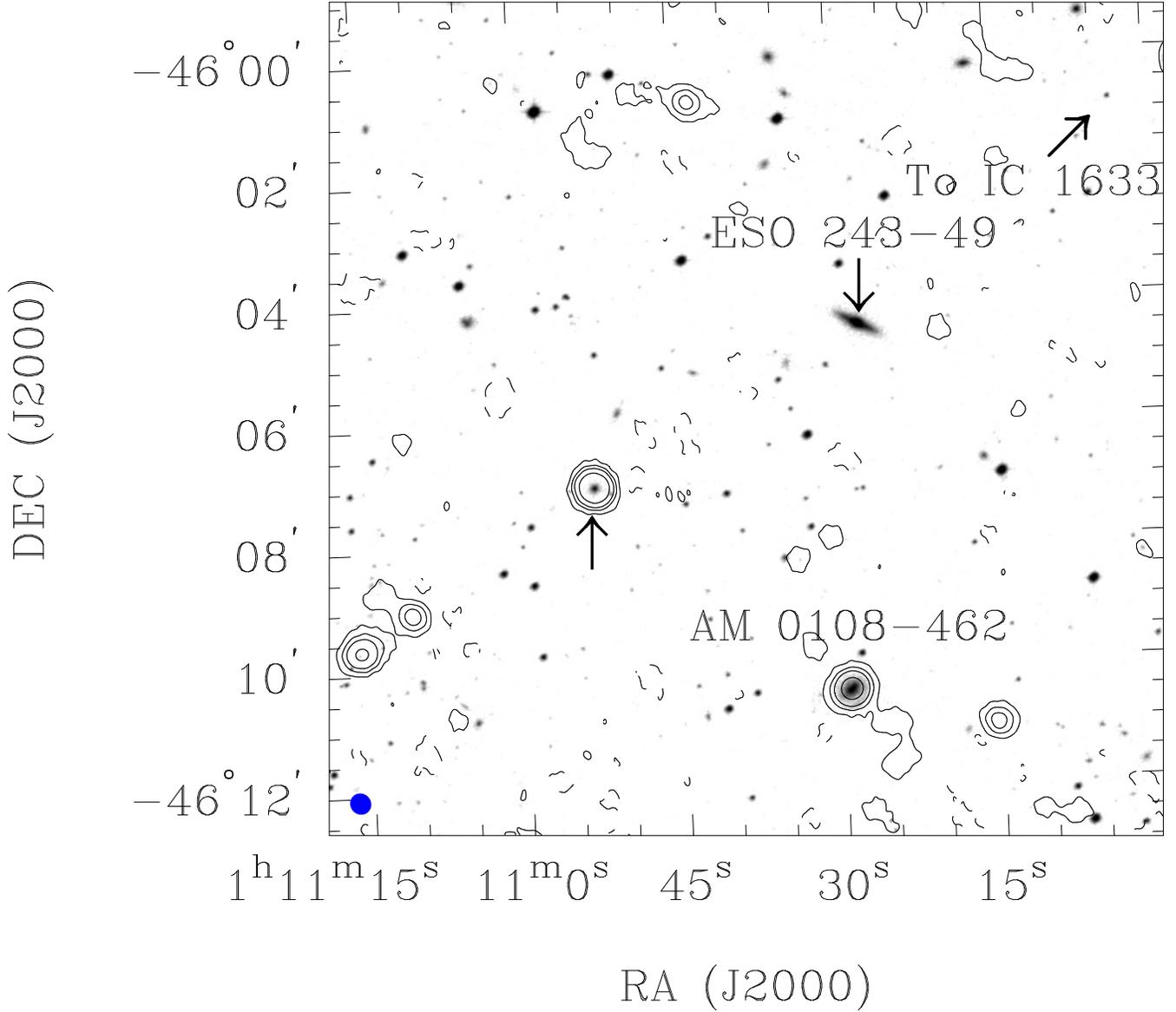}
 \caption{ATCA 20-cm radio map (contours) - overlaid on to an optical DSS-2 red image of the field of ESO 243-49. Contours are plotted in (-3, 3, 10, 20, 40)$\times \sigma$, where the rms noise $\sigma$=45 $\mu$Jy  beam$^{-1}$. Negative contours are dashed. The object marked with an upward arrow (with no label) is a background galaxy at a redshift of $\sim$0.1. The arrow in the top right corner points in the direction towards IC 1633 (not shown in this map), the central dominant galaxy of the Abell 2877 galaxy cluster. Synthesised beam (21.5 arcsec$\times$20.4 arcsec; position angle (PA) 37.8$^{\circ}$) is indicated at the bottom left.}
 \label{figRadioContinuum}
\end{figure*}

 We detect no continuum emission in or around ESO 243-49 down to a 5$\sigma$ upper limit of 225 $\mu$Jy  beam$^{-1}$ (see Figure \ref{figRadioContinuum}). The Phoenix Deep Survey conducted with ATCA at 1.4 GHz detected a continuum source PDF J011027.6-460427 at the centre of the galaxy with a flux density of 161$\pm$21 $\mu$Jy \citep{Hopkins2003}, which is at $\sim$3$\sigma$ level in our data and indistinguishable from noise spikes.

\par The radio continuum emission at 1.4 GHz $\left(L_{1.4\mathrm{GHz}}\right)$ from star-forming galaxies  is mainly synchrotron radiation produced by relativistic electrons. It can be used as an indication of the star formation rate ($SFR_{1.4\mathrm{GHz}}$) for stars of masses $\gtrsim 5 M_{\odot}$ that dominate this emission mechanism \citep{Condon1992}:

\begin{equation}
	SFR_{1.4\mathrm{GHz}}=\frac{L_{1.4\mathrm{GHz}}}{4.0\times 10^{21}\:\mathrm{W}\:\mathrm{Hz}^{-1}}\:M_{\odot} \:\mathrm{yr}^{-1}.
	\label{SFR_1.4GHz}
\end{equation}
Using this relation we estimated a star-formation rate of $<$0.06 $M_{\odot} \:\mathrm{yr}^{-1}$ in ESO 243-49 using the 5$\sigma$ upper limit on $L_{1.4\mathrm{GHz}}$ obtained in our data set \citep[compared to 0.04 $M_{\odot} \:\mathrm{yr}^{-1}$ derived from the flux density measured in the Phoenix Deep Survey,][]{Hopkins2003}.

\par No {H{\sc i} line emission is detected in or around ESO 243-49 to a 5$\sigma$ upper limit of 5.5 mJy  beam$^{-1}$. The equivalent masses of  H{\sc i} gas at 5$\sigma$ flux densities are given in Table \ref{tblHIupperLimits} (we assumed a top-hat line profile of various widths for this nearly edge-on \textit{S0a} galaxy and smoothed the spectra to various channel widths).

\begin{table}
\caption{$M_{\mathrm{H{\textsc i}}}$ upper limits for ESO 243-49.}
	\begin{center}
		\begin{tabular}{lll}
			\hline 
			\hline            
			Assumed line width&Channel width&$M_{\mathrm{H{\textsc i}}}$ ($10^8 M_{\odot}$)\\
			(km s$^{-1}$)&(km s$^{-1}$)&\\
			\hline
 			20&7&$<$2\\
			100&35&$<$5\\
			200&70&$<$9\\
			\hline  
		\end{tabular}
	\end{center}
\label{tblHIupperLimits}
\end{table}

\begin{figure}
 \includegraphics[width=0.48\textwidth]{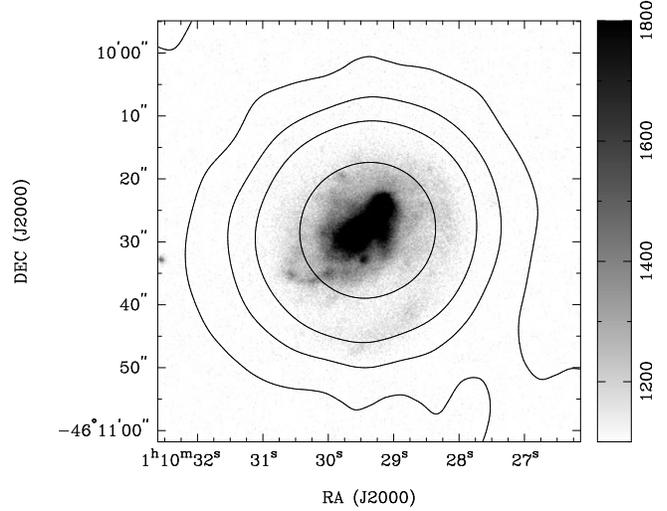}
 \caption{ATCA 20-cm radio map (contours) of the AM0108-462 galaxy - overlaid on to an optical \textit{RVB} image taken with the \textit{Magellan} 1 Telescope. Contours are (3, 10, 20, 40)$\times \sigma$, where the rms noise $\sigma$=45 $\mu$Jy  beam$^{-1}$. The colour bar for the optical image is in counts per pixel.}
 \label{figMagellanAM0108-462}
\end{figure}

\par We detect both continuum and {H{\sc i} line emission from AM 0108-462, a galaxy in the field of ESO 243-49, at a projected distance of $\sim$170 kpc away (see Figure \ref{figRadioContinuum}). No {H{\sc i} line emission is detected anywhere else in the field. Figure \ref{figMagellanAM0108-462} shows an optical image of this galaxy with the radio continuum emission contours overlaid. This peculiar flocculent \textit{Sb} galaxy shows knots of optical emission toward the nucleus, and emission lines in its optical spectrum indicate the galaxy has strong {H{\sc ii} regions \citep[][and references therein]{Hopkins2000}. Both galaxies lie at a distance of $\sim$95 Mpc \citep{Malumuth1992}, and belong to the Abell 2877 galaxy cluster \citep{Dressler1980}.

\par Figure \ref{figRadioContinuum} shows the continuum data for the radio source in AM 0108-462. Using the {\sc imfit} task we fitted a gaussian model component to the region of the continuum image containing the source in this galaxy. The source was extended and the total integrated flux density of the fitted source at 1.4 GHz, $S_{\mathrm{1.4 GHz}}$ was 5.25$\pm$0.19 mJy. At the same position (within the synthesised beam) Hopkins et al. \citeyearpar{Hopkins2000, Hopkins2003} reported an extended continuum source PDF J011029.4-461027 with integrated flux density of 4.201$\pm$0.092 mJy at 1.4 GHz. Their radio contours were similarly extended and centered on the optical nucleus. The flux density of the radio continuum source in AM 0108-462 in our data set is significantly higher than that in the Phoenix Deep Survey (PDF J011029.4-461027). The Phoenix survey used 6A, 6B and 6C array configurations and due to interference during the observations all the data from baselines shorter than a few hundred meters were flagged \citep[][and references therein]{Hopkins2003}. As a result the extended emission was resolved out and the flux density value reported was lower than that from our data set. Using equation \ref{SFR_1.4GHz} with $L_{1.4\mathrm{GHz}}$ obtained in our data set we estimated a star-formation rate of $\sim$1.5 $M_{\odot} \:\mathrm{yr}^{-1}$ in AM 0108-462.

\par For this radio source we also used 6-cm observations of AM 0108-462 conducted with ATCA \citep[for details of the observations and the data reduction procedures see][]{Cseh2014} to derive the radio spectral index, $\alpha_R$, between 5.5 and 1.4 GHz. The total integrated flux density of the source at 5.5 GHz was 1.72$\pm$0.15 mJy, and together with $S_{\mathrm{1.4 GHz}}$, derived above, $\alpha_R=-0.8$.

\par Figure \ref{figHIchannels} shows channel maps of  the H{\sc i} emission detected in AM 0108-462. The H{\sc i} emission covers a velocity range of $\sim$160 km s$^{-1}$, starting from $\sim$6,950 km s$^{-1}$. The H{\sc i} emission spectrum covering the velocity range of 1,000 km s$^{-1}$ centered on the emission is shown in Figure \ref{figHIspectrum}. Figure \ref{figHIIntensityVelocityFields}a shows a total intensity map of the H{\sc i} gas. We measured a total H{\sc i} flux density of 206 mJy km s$^{-1}$. At the distance of 95 Mpc this corresponds to an H{\sc i} mass of $\sim$$5\times10^8 M_{\odot}$. The gas is mostly located inside the galaxy and appears to form a rotating structure as demonstrated by the velocity gradient in Figure \ref{figHIIntensityVelocityFields}b. We measured the inclination angle of AM 0108-462 to be 49$^{\circ}$, similar the value of 53$^{\circ}$ reported in the Two Micron All Sky Survey Extended Source Catalogue \citep{Skrutskie2006}. The inclination angle along with the apparent velocity gradient of H{\sc i} emission argues against this galaxy being face-on as previously reported by Hopkins \citeyearpar{Hopkins2000}.

\par We also detect  a narrow feature outside of the galaxy appearing only in five channels (this narrow feature is centered at the radio velocity of 7,094 km s$^{-1}$ in Figure \ref{figHIspectrum}), with a total H{\sc i} mass of $\sim$$8\times10^7 M_{\odot}$. The velocity range of this H{\sc i} clump is within that of the main body of AM 0108-462 as can be seen in Figure \ref{figHIIntensityVelocityFields}b and also from the H{\sc i} spectrum (Figure \ref{figHIspectrum}). There is no obvious optical emission $\sim$21 kpc from the centre of the galaxy, at the position of the H{\sc i} structure (see Figure \ref{figHIchannelMagellanSwiftUVOT}a), but the \textit{Swift} Ultra-Violet/Optical Telescope (UVOT) detects ultraviolet emission (using the UVW2 filter with the central wavelength of 188 nm) as can be seen in Figure \ref{figHIchannelMagellanSwiftUVOT}b. Having detected the H{\sc i} line emission in AM 0108-462, we reduced and analysed X-ray and optical data of this galaxy, as described next.

\subsection{X-ray data of AM 0108-462}

\par We analysed the archival data of an X-ray source 3XMM J011029.5-461026 (hereafter, referred to as XMM J0110-46) detected by \textit{XMM-Newton} in the centre of AM 0108-462. The field of AM 0108-462 has been observed several times, the observation with the highest signal-to-noise ratio data for this X-ray source was conducted on May 14 2010. The exposure time was 106.3 ks with the European Photon Imaging Camera (EPIC; pn, MOS1 and MOS2 cameras) covering the energy range from 0.2 to 12 keV. AM 0108-462 was outside of the field of view of the pn camera, hence only the data from the MOS cameras were available. The parameters of this observation are listed in Table \ref{tblXMM-NewtonObs}.

\renewcommand{\arraystretch}{1.2}
\begin{table*}
\caption{Details of the {\it XMM-Newton} observation of XMM J0110-46 in AM 0108-462.}
	\begin{center}
		\begin{tabularx}{\textwidth}{XXlXXXXl}
			\hline 
			\hline                       
  			R.A.& Dec.& Positional Accuracy &Telescope &Instrument & Date & Observation& Exposure\\
			 (J2000)& (J2000)&(at 3$\sigma$ level)& & & & ID&time (ks)\\
						\hline
 			01$^{h} $10$^{m} $29.58$^{s}$ &-46$^{\circ} $10$'$ 26.8$''$&2.07 arcsec& {\it XMM-Newton} & MOS1\&2 &2010-05-14 & 0655510201 & 106.3  \\
 			 \hline  
		\end{tabularx}
	\end{center}
\label{tblXMM-NewtonObs}
\end{table*}

\par The Science Analysis Software (SAS) version 13.5 software\footnote{\url{http://xmm.esa.int/sas/}} was used to process the observational data files, following the procedure outlined in Servillat et al. \citeyearpar{Servillat2011}. The data were reduced using the \texttt{emproc} script with the most recent calibration data files. For the extraction of spectra good time interval (GTI) filtering was applied to remove intervals with high background rates (above 0.45 counts $\mathrm{s^{-1}}$ for both MOS cameras) due to such flares to maximise the signal-to-noise ratio. The source spectra were extracted using circular regions centred on the source position with radii of 20 arcsec. Background spectra were extracted from nearby source-free regions of the field with areas twice the source extraction region areas.

\par The spectra extracted from this observation were fitted with power law (\textit{pow}) and thermal plasma (\textit{apec}) models using XSPEC\footnote{\url{http://heasarc.nasa.gov/docs/xanadu/xspec}} v12.8.1g \citep{Arnaud1999}. For each model photoelectric absorption was accounted for by using the \textit{phabs} component in XSPEC and the Wilms abundances \citep{Wilms2000}. The weighted average Galactic H{\sc i} absorption in the direction of XMM J0110-46 as measured by the Leiden/Argenitine/Bonn (LAB) survey of Galactic H{\sc i} is $1.8\times10^{20}\:\mathrm{cm}^{-2}$ \citep{Kalberla2005}. We restricted the minimum value for the $n_H$ parameter (absorption on the line of sight) to the Galactic value. Acceptable fits were obtained for both models with all the relevant parameters listed in Table \ref{tblSpectralFitParsXMMJ0110-46}.

\par Whilst the Astrophysical Plasma Emission Code (APEC) thermal emission model produced an acceptable fit, the plasma temperature of $\sim$10 keV is too high for a star formation region \citep{Mineo2012}. The power law model fit is consistent with inverse Compton emission from an accreting black hole. The photon index value for the model is typical of that for an active galactic nucleus (AGN) \citep[e.g.,][]{Page2005}. Assuming the X-ray source is in AM 0108-462, its luminosity of $\sim$$10^{40}\mathrm{\:erg\:s^{-1}}$ is typical of a low-luminosity AGN. The spectral fit for the power law model along with the residuals is presented in Figure \ref{XMMJ0110-46Spectra}.

\begin{figure*}
 \includegraphics[width=1\textwidth]{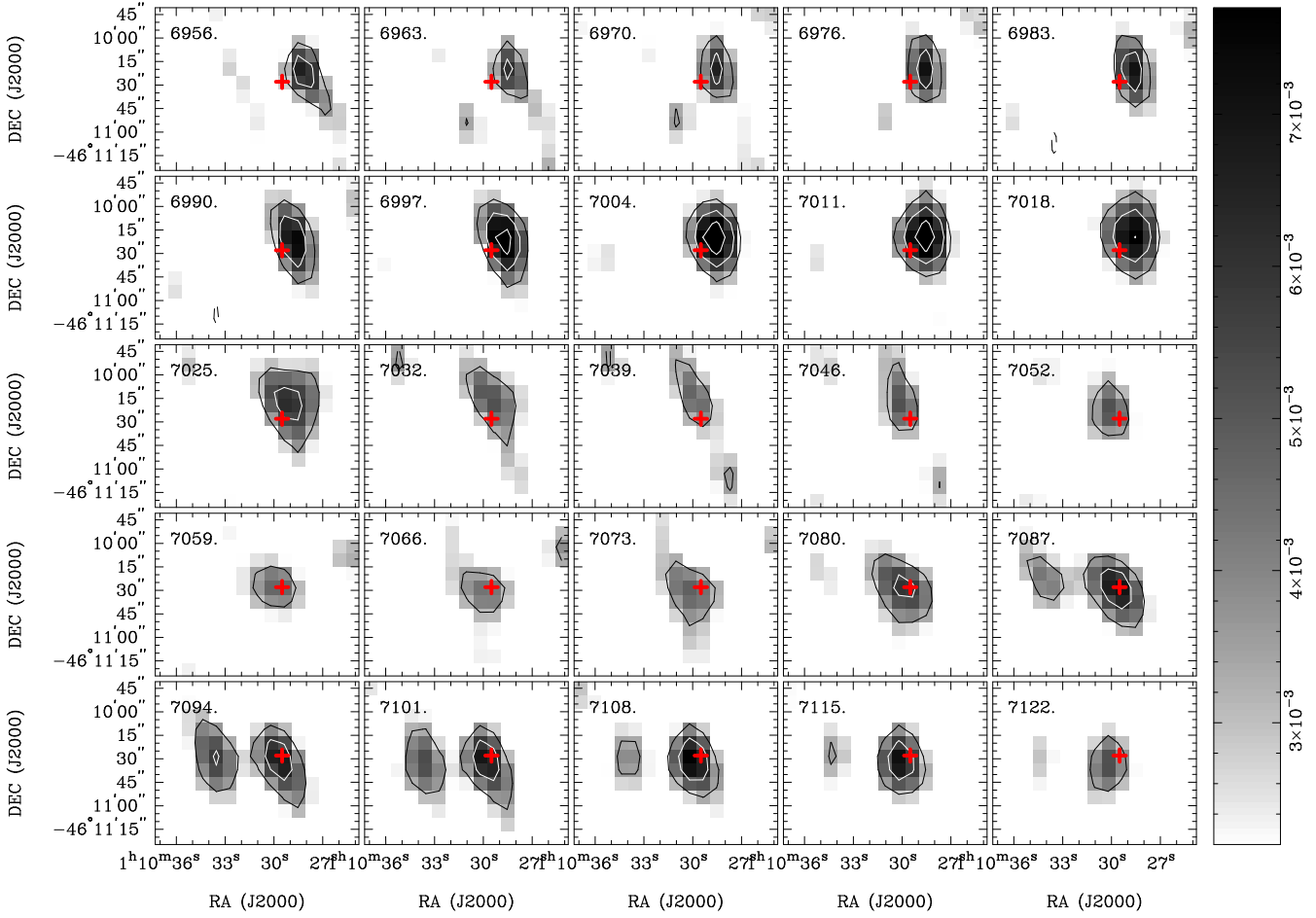}
 \caption{H{\sc i} channel maps in AM 0108-462. Contour levels are at (-3, 3, 5, 7)$\times \sigma$, where the rms noise $\sigma$=1.1mJy  beam$^{-1}$. Negative contours are dashed. The colour bar is in Jy  beam$^{-1}$. The velocity labels in the top left corner are in km s$^{-1}$. The cross marks the position of the centre of the galaxy at 01$^{h}$ 10$^{m}$ 29.5$^{s}$ -46$^{\circ}$ 10$'$ 28$''$ \citep[J2000;][]{Skrutskie2003}.}
 \label{figHIchannels}
\end{figure*}

\renewcommand{\arraystretch}{1.2}
\begin{table*}
\caption{Spectral models applied to the {\it XMM-Newton} spectra of XMM J0110-46. All errors are quoted at 90\% confidence level.}
\begin{threeparttable}
	\begin{center}
		\begin{tabularx}{\textwidth}{lXXXXc}
			\hline 
			\hline                       
  			Model& $n_H$\tnote{1} & $\Gamma/kT$\tnote{2}& $F_\mathrm{{\left[0.2-10\:keV\right]}}$\tnote{3} &$L_\mathrm{{\left[0.2-10\:keV\right]}}$\tnote{4} &${\chi^2}/$degrees of freedom\\
			\hline
			{\it pow} & $1.8-4.0$ & $1.6\pm0.2$  & $4.9^{+3.0}_{-1.7}$ & $5.3^{+3.2}_{-1.8}$ & 39/38\\
			{\it apec} & $1.8-2.3$  & $8.7^{+17.4}_{-4.4}$ & $4.7^{+1.5}_{-1.2}$ & $5.1^{+1.6}_{-1.3}$ & 42/37\\
 			\hline  
		\end{tabularx}
		\begin{tablenotes}
            		\item[1] Absorption along the line of sight (in $10^{20}\:\mathrm{atom\:cm}^{-2}$);
			\item[2] Photon index (dimensionless) for the \textit{pow} model, plasma temperature (in keV) for the \textit{apec} model;
			\item[3] Absorbed flux in the 0.2$-$10 keV energy range for the model $\mathrm{\left(in\:10^{-14}\:erg\:cm^{-2}\:s^{-1}\right)}$;
			\item[4] Absorbed luminosity in the 0.2$-$10 keV energy range $\mathrm{\left(in\:10^{40}\:erg\:s^{-1}\right)}$.
		\end{tablenotes}
	\end{center}
\label{tblSpectralFitParsXMMJ0110-46}
\end{threeparttable}
\end{table*}

\subsection{Optical data of AM 0108-462}\label{secOpticalData}

\par The field of AM 0108-462 was observed with IMACS and PANIC at the Magellan 6.5 m telescopes in August 2010 in different optical bands (\textit{B}, \textit{V}, \textit{R} with IMACS) and in the near-infrared band (\textit{J} band with PANIC). Spectroscopic observations along the major axis of the galaxy (slit spectroscopy mode with IMACS) were performed at the same time. The images presented in Figure \ref{figAllbands} show the complexity of the structure in AM 0108-462, which appears to be asymmetric and with two bright knots located in its central region along the major axis of this galaxy in the ultraviolet and possibly in the optical $B$ images. However, moving through the longer wavelength optical to near-infrared bands one of the knots starts to look like an arc or a spiral arm, especially in the $J$ image, where it appears as an elongated single region (Figure \ref{figAllbands}).

\par The position of the slit and the 2D spectrum around H$\alpha$ are shown in Figure \ref{figSpecrv}a. Narrow emission lines (FWHM of 8\AA, the resolution of the instrument) are clearly detected. In a standard Baldwin, Phillips and Terlevich (BPT) diagram \citep{Baldwin1981}, all the emitting regions are classified as H{\sc ii} regions, where intense star formation is occurring. In contrast, none of the typical markers of AGN-like activity were detected. Along the galaxy, we extracted a spectrum for 12 sub-apertures. We measured the radial velocities of the H{\sc i}-emitting gas along the slit and they are similar to those of the H$\alpha$-emitting gas (see Figure \ref{figSpecrv}b). The region of H{\sc i} emission extends beyond that of the optical and near-infrared emission as indicated by the white bold ellipse (marking the size of the galaxy in \textit{K} band) and line (marking the size of the slit) in Figure \ref{figHIIntensityVelocityFields}b.

\section{Discussion}\label{secDiscussion}

\subsection{The lenticular galaxy ESO 243-49}

\par To interpret the H{\sc i} non-detection of the galaxy ESO 243-49 we first estimated the expected H{\sc i} mass of ESO 243-49 based on its morphological type and blue luminosity. Whilst $\sim$$10^9 M_{\odot}$ of H{\sc i} gas has been detected around a number of \textit{S0} galaxies \citep[e.g.,][]{Emonts2008,Struve2010,Serra2012}, the ratio of a galaxy's H{\sc i} mass (in $M_{\odot}$) to its blue luminosity (in $L_{\odot}$) is a common diagnostic for \textit{S0a} galaxies and a typical value is $\sim$0.04 \citep{Wardle1986}. Given ESO 243-49's blue luminosity of $\sim$$1\times10^{10} L_{\odot}$ \citep{Lauberts1989}, to achieve a ratio of 0.04 an H{\sc i} mass of $\sim$$6\times10^8 M_{\odot}$ is required. This estimate requires further adjustment for the environment of ESO 243-49 since it is at a projected distance of $\sim$0.3 Mpc away from IC 1633, the central dominant galaxy of its galaxy cluster. Serra et al. \citeyearpar{Serra2012} reported that the galaxies inside $\sim$1 Mpc from the centre of the Virgo galaxy cluster were significantly H{\sc i}-deficient compared to the galaxies outside the cluster. A common measure of H{\sc i} deficiency is the H{\sc i} deficiency parameter, defined as
\begin{equation}
	\mathrm{Def_{H{\textsc i}}}=log(M_\mathrm{{H{\textsc i} \: exp}})-log(M_\mathrm{{H{\textsc i}}}),
\end{equation}
where $M_\mathrm{{H{\textsc i} \: exp}}$ is the expected H{\sc i} mass, and $M_\mathrm{{H{\textsc i}}}$ is the observed H{\sc i} mass. A galaxy is considered H{\sc i}-deficient if Def$_\mathrm{{H{\textsc i}}}>0.5$ \citep[e.g.,][]{Cortese2011}. However, just how H{\sc i}-deficient an early-type galaxy near the cluster centre gets is not clear. Serra et al. \citeyearpar{Serra2012} report that in their sample the H{\sc i} detection rate is 4$\pm$4\% for early-type galaxies close to the centre of the Virgo galaxy cluster as compared to the detection rate of 23$\pm$13\% outside 1 Mpc from the cluster centre. Assuming the minimum value for Def$_\mathrm{{H{\textsc i}}}=0.5$ for ESO 243-49, we estimated its H{\sc i} mass to be at most $\sim$$2\times10^8 M_{\odot}$. This amount of H{\sc i} gas cannot be detected in our line data set with a line width of $>$50 km  s$^{-1}$ typical of edge-on galaxies (see Table \ref{tblHIupperLimits}; at 5$\sigma$, assuming a top-hat profile). Therefore ESO 243-49 is not a typical \textit{S0a} galaxy in its H{\sc i} content.

\par Assuming that HLX-1 is indeed a remnant of a dwarf satellite galaxy of ESO 243-49, we estimated the H{\sc i} content of the dwarf using the non-detection limits in our line data set. The H{\sc i} masses of dwarf galaxies vary depending on their morphological type. The most H{\sc i}-rich dwarf galaxies are the dwarf irregulars, including the blue compact dwarfs that have $\lesssim$$10^9 M_{\odot}$ of H{\sc i} gas. Dwarf ellipticals have  $\lesssim$$10^8 M_{\odot}$ of H{\sc i} gas, and the most H{\sc i}-poor are the dwarf spheroidals with $\lesssim$$10^5 M_{\odot}$ of H{\sc i} gas \citep[for a review see][and references therein]{Grebel2001}. Since in our line data set we cannot detect $\lesssim$$10^8 M_{\odot}$ of H{\sc i} gas, we considered further only the H{\sc i}-rich dwarfs. However, as was the case with ESO 243-49, the environment also needs to be taken into account when estimating the H{\sc i} mass.

\par Lee et al. \citeyearpar{Lee2003} constructed a sample of dwarf irregulars in the Virgo galaxy cluster to examine the impact of the cluster environment on galaxy H{\sc i} content. They found that five Virgo dwarfs closest to the cluster centre were gas-deficient by a factor of between 8 and 30 (ranging in the H{\sc i} mass between 2 and 5 $\times$ 10$^7$ $M_{\odot}$) compared to a control sample of the field dwarf irregulars. VCC 1217, another dwarf irregular in the Virgo cluster, is $\sim$0.3 Mpc away (in projected distance) from the cluster dominant galaxy (M 87), the same distance as ESO 243-49 is from its cluster dominant galaxy. VCC 1217 has no H{\sc i} gas detected down to 8 $\times$ 10$^6$ $M_{\odot}$ in the main body of the galaxy \citep{Chung2009}. The outer tail of this galaxy formed via ram pressure stripping contains 4 $\times$ 10$^7$ $M_{\odot}$ of H{\sc i} gas, only $\sim$6\% of the expected H{\sc i} content for a late-type dwarf with the mass of VCC 1217 \citep{Kenney2014}. In our line data set we cannot detect this little H{\sc i} gas even in the case of a narrow line width of 20 km s$^{-1}$ (see Table \ref{tblHIupperLimits}; at 5$\sigma$, assuming a top-hat profile).

\par The H{\sc i} non-detection in ESO 243-49 including the dwarf satellite galaxy is therefore likely a result of these galaxies being significantly poorer in H{\sc i} than expected for such galaxies in the field due to the location of ESO 243-49 close to the centre of its galaxy cluster. The H{\sc i} emission upper limits in our data rule out the possibility of HLX-1 being a remnant of a gas-rich dwarf satellite of ESO 243-49 but are consistent with a dwarf having a moderate or poor reservoir of H{\sc i} gas. To investigate further more sensitive H{\sc i} observations of ESO 243-49 and the field around it are required.

\begin{figure}
 \includegraphics[width=0.49\textwidth]{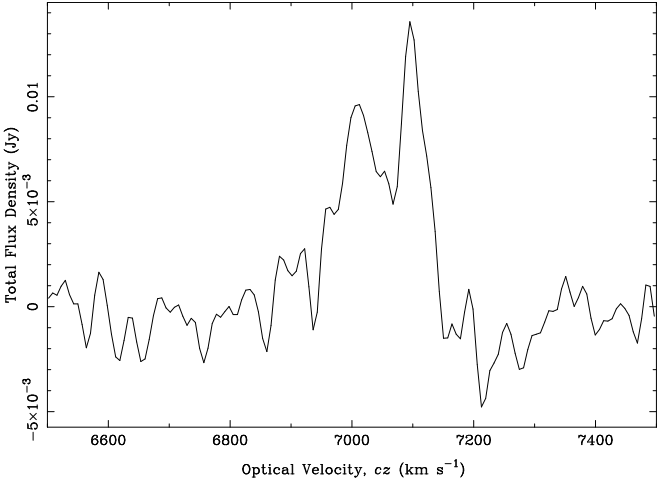}
 \caption{The H{\sc i} emission spectrum of AM 0108-462.}
 \label{figHIspectrum}
\end{figure}

\begin{figure*}
\begin{center}$
\begin{array}{cc}
\includegraphics[width=0.5\textwidth]{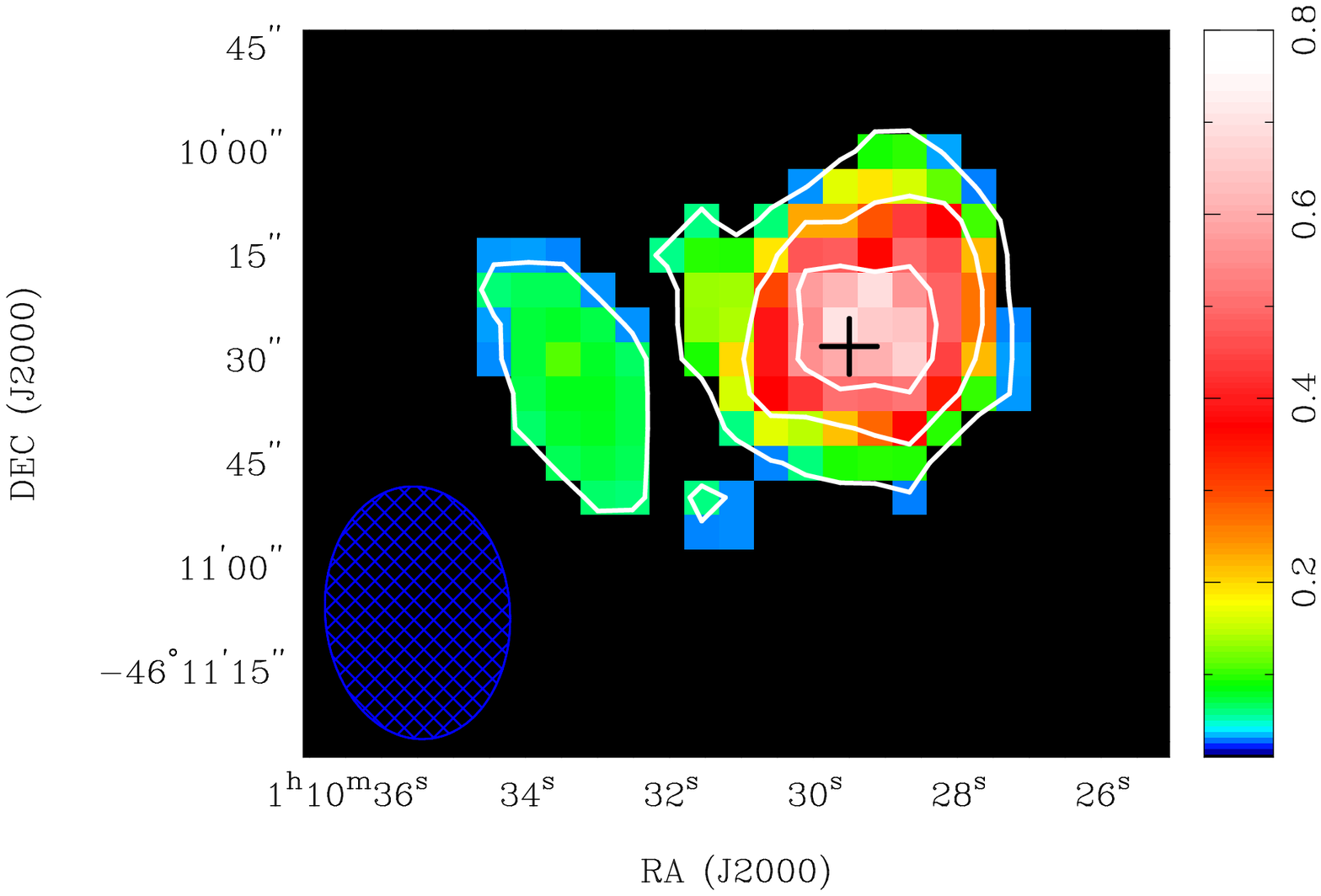} &
\includegraphics[width=0.5\textwidth]{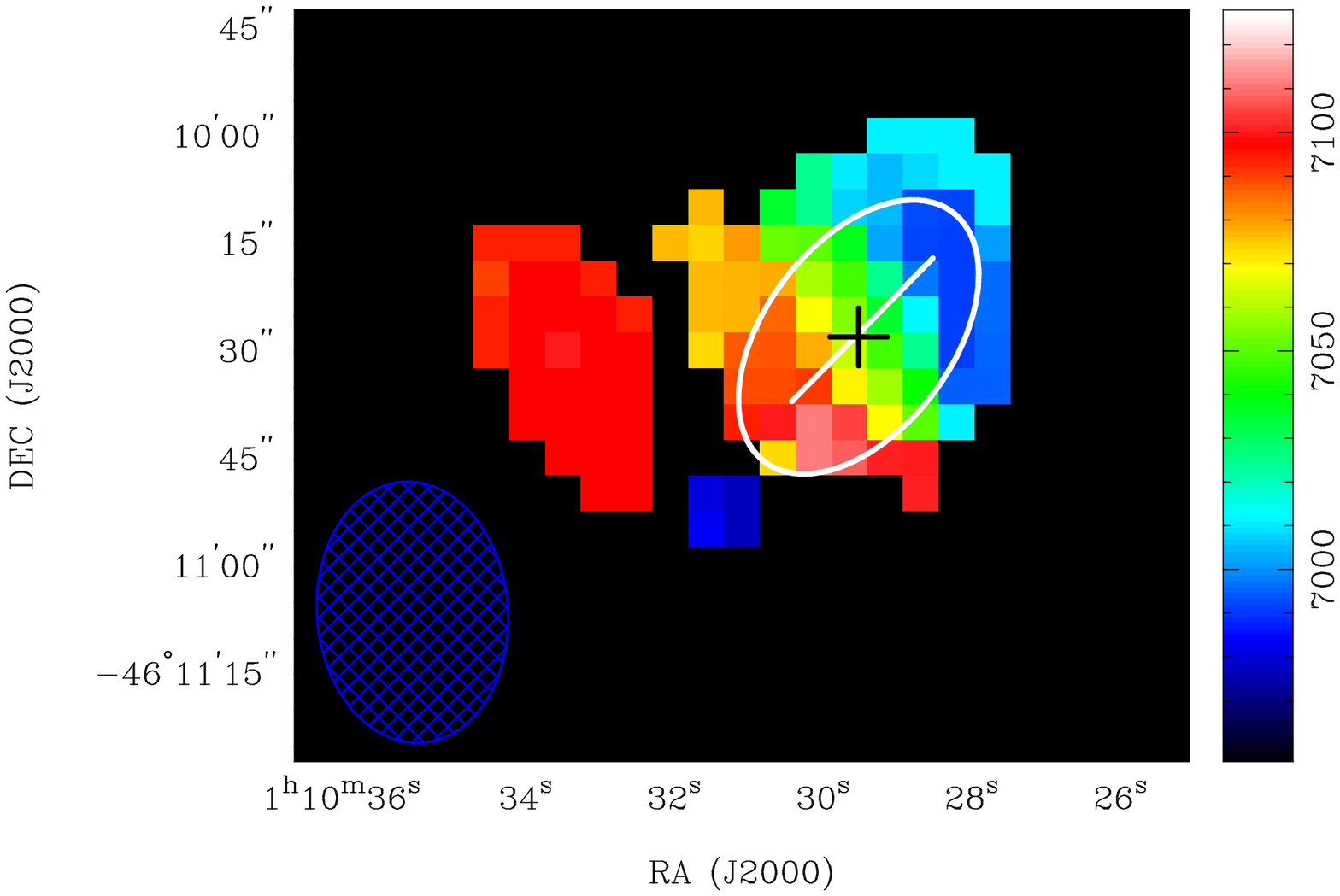}\\
{\mbox{\bf a}} &
{\mbox{\bf b}}
\end{array}$
\end{center}
\caption {\textbf{a}: The integrated H{\sc i} intensity distribution in AM 0108-462 with contour levels at 0.04, 0.24 and 0.42 Jy beam$^{-1}$ km s$^{-1}$. \textbf{b}: The H{\sc i} velocity field in AM 0108-462. The velocity units are in km s$^{-1}$. The white ellipse marks the size of the galaxy in \textit{K} band; the line represents the position of the slit for the spectroscopic observations described in section \ref{secOpticalData}; the cross represents the centre of the galaxy. The blue hatched ellipse (bottom left) represents the size of the synthesised beam (37.2 arcsec$\times$27.0 arcsec; PA of 7.1$^{\circ}$).}
\label{figHIIntensityVelocityFields}
\end{figure*}

\subsection{The spiral galaxy AM 0108-462}

\par We estimated the expected H{\sc i} mass of AM 0108-462 based on its absolute magnitude in several optical and near-infrared wavebands using scaling relations of the form
\begin{equation}
	log(M_\mathrm{{H{\textsc i}}})=\alpha+\beta M_x,
	\label{eqnHImassFromMagnitude}
\end{equation}
where $M_x$ is the absolute magnitude in the various bands and $\alpha$ and $\beta$ are the parameters of the relation \citep[][see Table 2 for the parameter values and the wavebands]{Denes2014}. We derived the absolute magnitudes in several wavebands using published apparent magnitude values to estimate $log(M_\mathrm{{H{\textsc i}}})=9.6\pm0.4$.

\par The expected values exceed the H{\sc i} mass we measured in our line data set. However, as was the case with ESO 243-49, AM 0108-462's environment needs to be considered. Like ESO 243-49, AM 0108-462 too is close to the cluster dominant galaxy ($\sim$0.4 Mpc away in projected distance). To quantify the effect of the environment, we used the relation between galaxy stellar ($M_{*}$) and H{\sc i} mass for galaxies in the Virgo cluster \citep{Cortese2011}. Cortese et al. \citeyearpar{Cortese2011} found the fraction of the stellar mass to the H{\sc i} mass, $log(M_\mathrm{{H{\textsc i}}}/M_{*})\sim-2$ for the Virgo galaxies with the same stellar mass as AM 0108-462 \citep[$\sim$6$\times 10^{10} M_{\odot}$ estimated from its \textit{K}-band luminosity;][]{Baldry2003}. AM 0108-462's measured H{\sc i} mass of $\sim$5$\times 10^{8} M_{\odot}$ in our line data set gives the same value for $log(M_\mathrm{{H{\textsc i}}}/M_{*})\sim-2$, showing that the cluster environment has a similar effect on the H{\sc i} gas content of AM 0108-462 as it does for the galaxies in the Virgo galaxy cluster of comparable stellar mass. Thus, the H{\sc i} detection in AM 0108-462 acts as an excellent probe of the cluster environment around ESO 243-49, providing evidence for the H{\sc i} gas reservoir depletion, explaining the non-detection in ESO 243-49.

\par The H{\sc i} clump east of the main body of AM 0108-462 is likely associated with the galaxy since its H{\sc i} gas velocity range coincides with that of AM 0108-462. Its association with the galaxy along with the presence of ultraviolet emission and the lack of significant optical emission suggests that AM 0108-462 might be an extended ultraviolet disc (XUV disc) galaxy \citep[][and references therein]{Goddard2011}. These discs are quite common in spiral galaxies and ultraviolet emission can extend well beyond the optical edge. Many of the ultraviolet knots in XUV discs correspond to structures in H{\sc i} distribution and have no obvious or very faint optical counterparts \citep[][and references therein]{Goddard2011}.

\par We estimated AM 0108-462's 24-$\mu$m infrared luminosity, $L_{\mathrm{24 \mu m}}$ to be 3.7$\times10^{43}$ erg s$^{-1}$ from the IRAS catalog, which converts into a star formation rate of $\sim$8 $M_{\odot}$ yr$^{-1}$ using the \citet{Rieke2009} relation. Both the star formation rate and the morphology of AM 0108-462 indicate that it may be in the coalescence phase of a merger \citep[phase (d) in Figure 1 of][]{Hopkins2008}, right before an intense episode of star formation ($\sim$100 $M_{\odot}$ yr$^{-1}$) with enhanced emission ($L_{\mathrm{IR}}>10^{12}$$L_{\odot}$). The arms seen around the galaxy (in Figure \ref{figAllbands}) could correspond to projected tidal tails, as generally observed during galaxy mergers \citep[e.g.,][]{Hopkins2008,Cox2006}. Assuming the merger scenario, the lack of evidence in the optical spectrum of broad lines typical of AGN \citep[for example, H$\alpha$ line in the spectrum of AM 0108-462 reported in][]{Hopkins2000} and the X-ray emission from the galaxy centre can be explained by a low-luminosity AGN that either has only recently turned on, or has a small sphere of influence (and thus a small SMBH mass).

\par The far-infrared-radio flux ratio parameter $q$ \citep[equation 2,][]{Mauch2007} is often used as a diagnostic to distinguish between star-forming galaxies and radio-loud AGN. Using far-infrared flux densities $S_{\mathrm{60 \mu m}}=0.438$ Jy and $S_{\mathrm{100 \mu m}}=1.229$ Jy from the IRAS catalogue, and the radio flux density $S_{\mathrm{1.4 GHz}}=5.25$ mJy from our data set, we estimate $q=2.2$, suggesting AM 0108-462 is a star-forming galaxy \citep[][and references therein]{Mauch2007}. Another such diagnostic is the far-infrared spectral index between 60 and 25 $\mu$m, $\alpha_{\mathrm{FIR}}=$ log$(S_{\mathrm{25 \mu m}}/S_{\mathrm{60 \mu m}})$/log(60/25). $\alpha_{\mathrm{FIR}}$ is a measure of the temperature of the far-infrared emitting dust. Star-forming galaxies have cooler dust emission because of their extended star and gas distributions, and therefore have values of $\alpha_{\mathrm{FIR}}<-1.5$. The central engines of AGN heat the dust to warmer temperatures and have values of $\alpha_{\mathrm{FIR}}>-1.5$ \citep[][and references therein]{Mauch2007}. Using far-infrared flux densities $S_{\mathrm{25 \mu m}}=0.285$ Jy and $S_{\mathrm{60 \mu m}}$ from the IRAS catalogue, we estimate $\alpha_{\mathrm{FIR}}=-0.5$, typical of radio-loud AGN.

\begin{figure*}
\begin{center}$
\begin{array}{cc}
\includegraphics[width=0.5\textwidth]{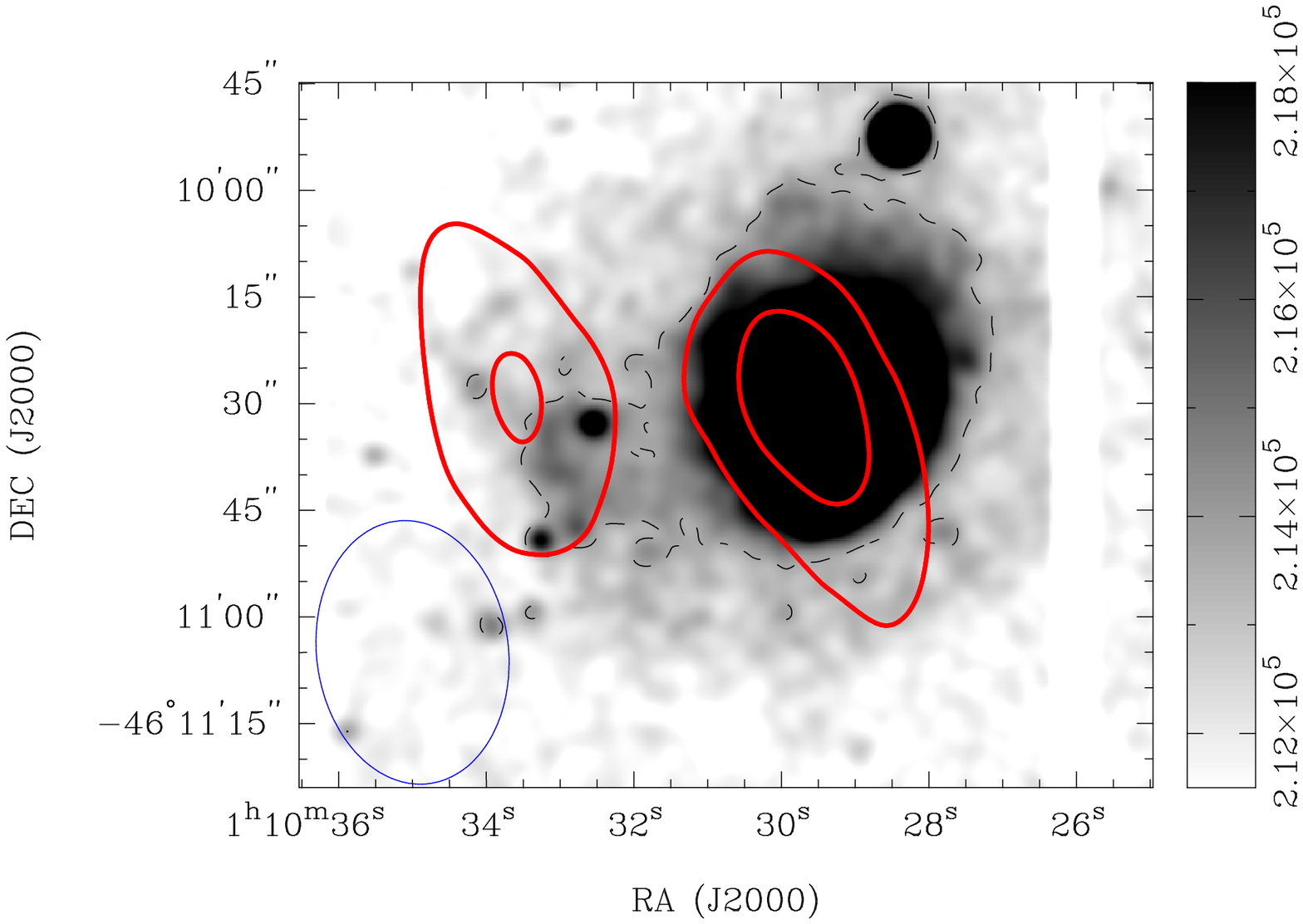} &
\includegraphics[width=0.5\textwidth]{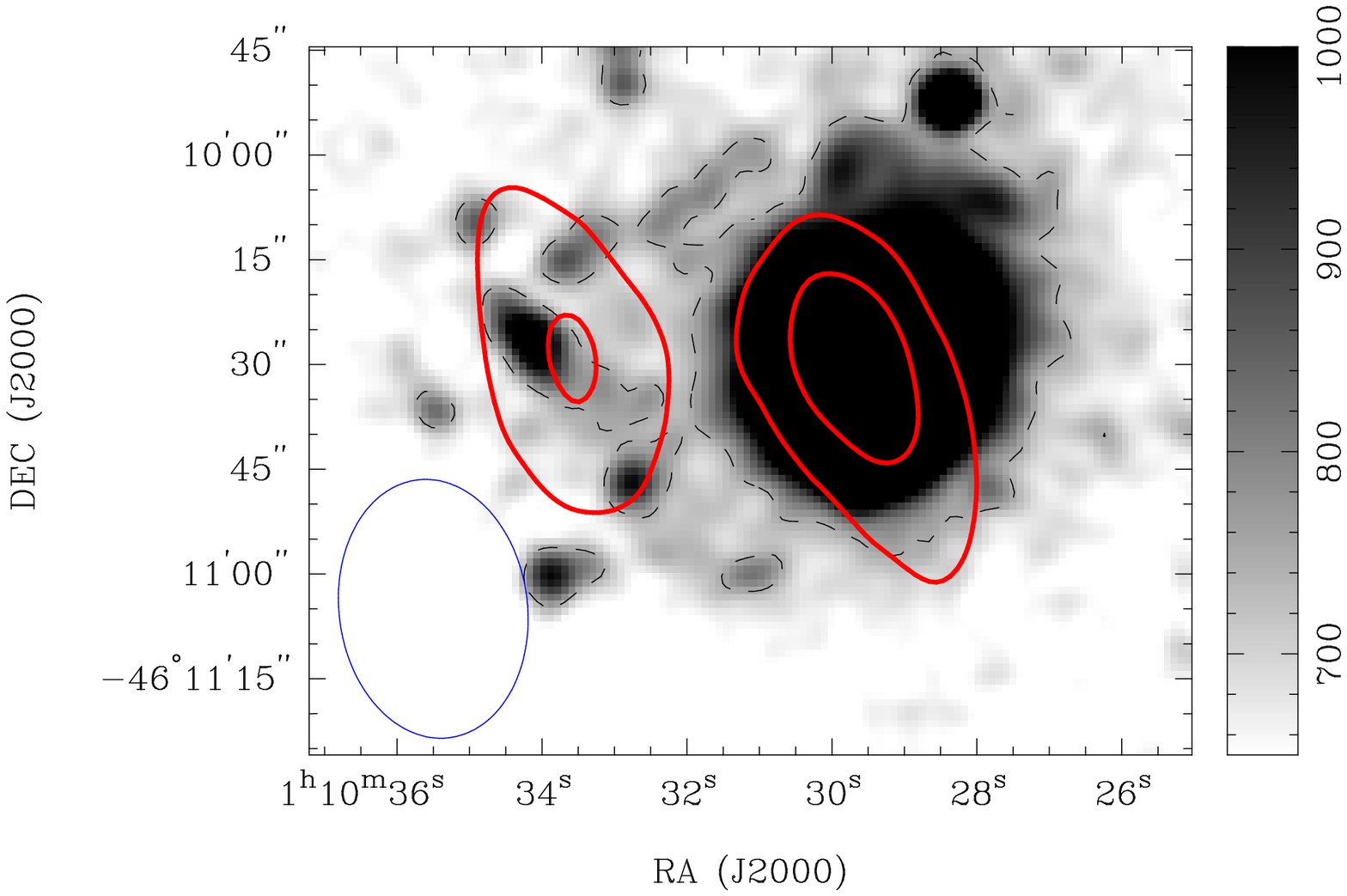}\\
{\mbox{\bf a}} &
{\mbox{\bf b}}
\end{array}$
\end{center}
\caption {\textbf{a}: A channel map of H{\sc i} emission in AM 0108-462 at velocity of 7,094 km s$^{-1}$ - overlaid on to a \textit{Magellan} 1 Telescope image (using the \textit{R}-band filter). This particular velocity value is chosen to display the association of the H{\sc i} clump with the H{\sc i} emission from the main body of AM 0108-462. Optical contours are dashed. H{\sc i} contours are bold red and the levels are at (3, 5)$\times \sigma$, where the rms noise $\sigma$=1.1mJy  beam$^{-1}$. The colour bar is in counts per pixel. \textbf{b}: A channel map of H{\sc i} emission in AM 0108-462 at velocity of 7,094 km s$^{-1}$ - overlaid on to a \textit{Swift} UVOT image (using the UVW2 filter). Ultraviolet contours are dashed.}
\label{figHIchannelMagellanSwiftUVOT}
\end{figure*}

\begin{figure}
  \centering
      \includegraphics[width=0.49\textwidth]{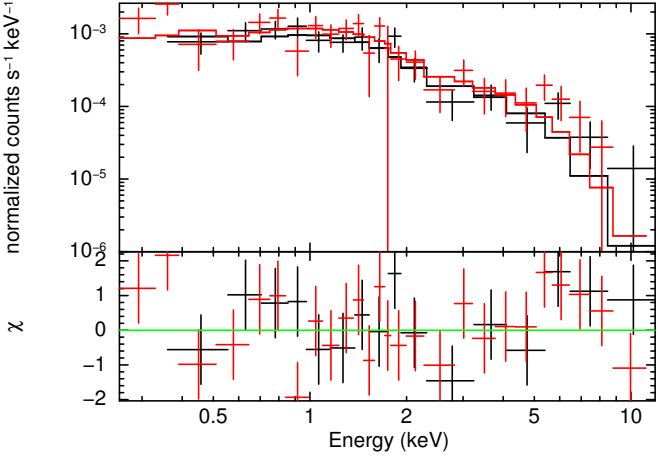}
  \caption{{\it XMM-Newton} unfolded EPIC spectra (black $=$ MOS1, red $=$ MOS2) of XMM J0110-46 fitted with the absorbed power law model. The bottom panel shows the fit residuals.}
  \label{XMMJ0110-46Spectra}
\end{figure}


\begin{figure*}
\includegraphics[width=1\textwidth]{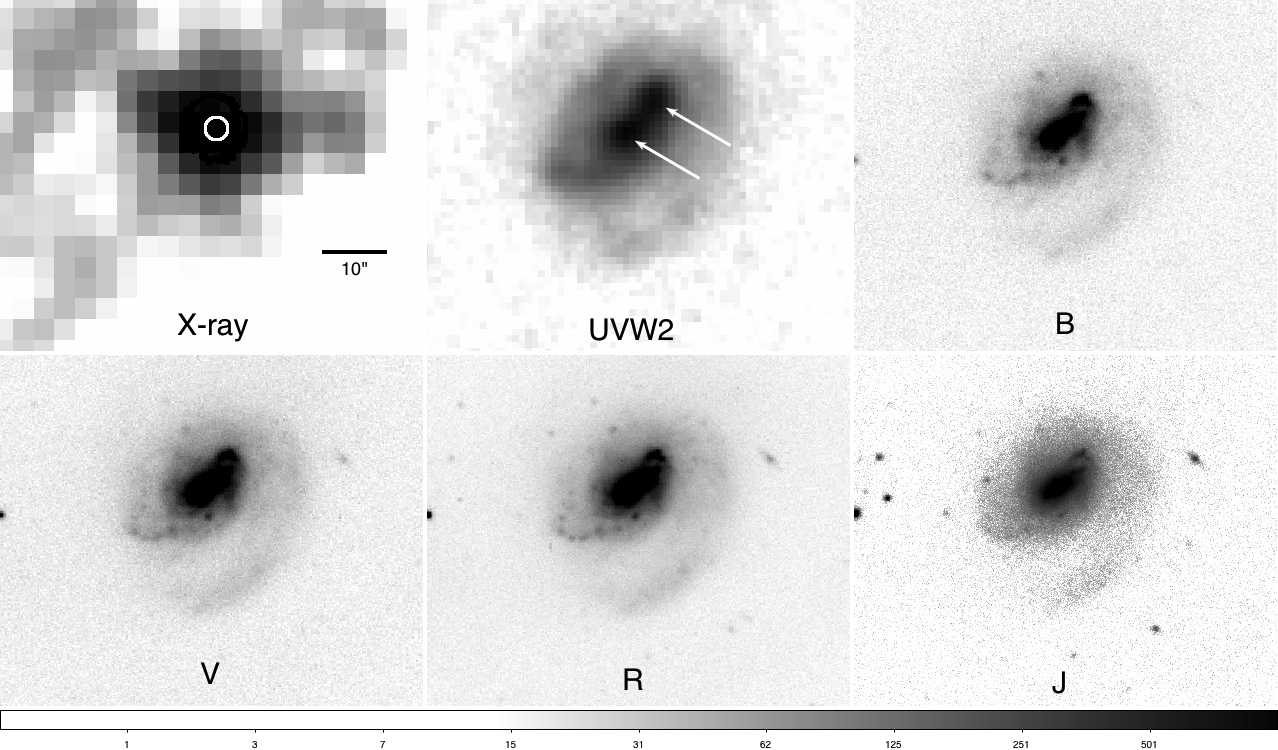}
\caption {The galaxy AM 0108-462 in different bands. The X-ray image was obtained with \textit{XMM-Newton}, the ultraviolet image with \textit{Swift} UVOT (using the UVW2 filter; white arrows indicate the two bright knots of ultraviolet emission in its central region), the \textit{B}, \textit{V}, and \textit{R} images with IMACS at the Magellan telescopes, and the \textit{J} image with PANIC at the Magellan telescopes. The 2.07 arcsec 3$\sigma$ error circle of the X-ray position is indicated in the first image. Each image is 1 arcmin large, covering the same region of AM 0108-462, with North up and East left.}
\label{figAllbands}
\end{figure*}

\begin{figure*}
\begin{center}$
\begin{array}{cc}
\includegraphics[width=.41\textwidth]{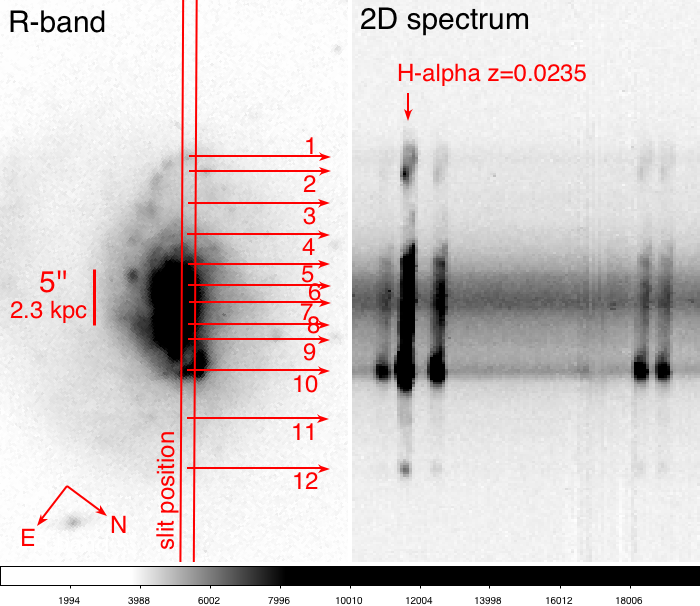} &
\includegraphics[width=.58\textwidth]{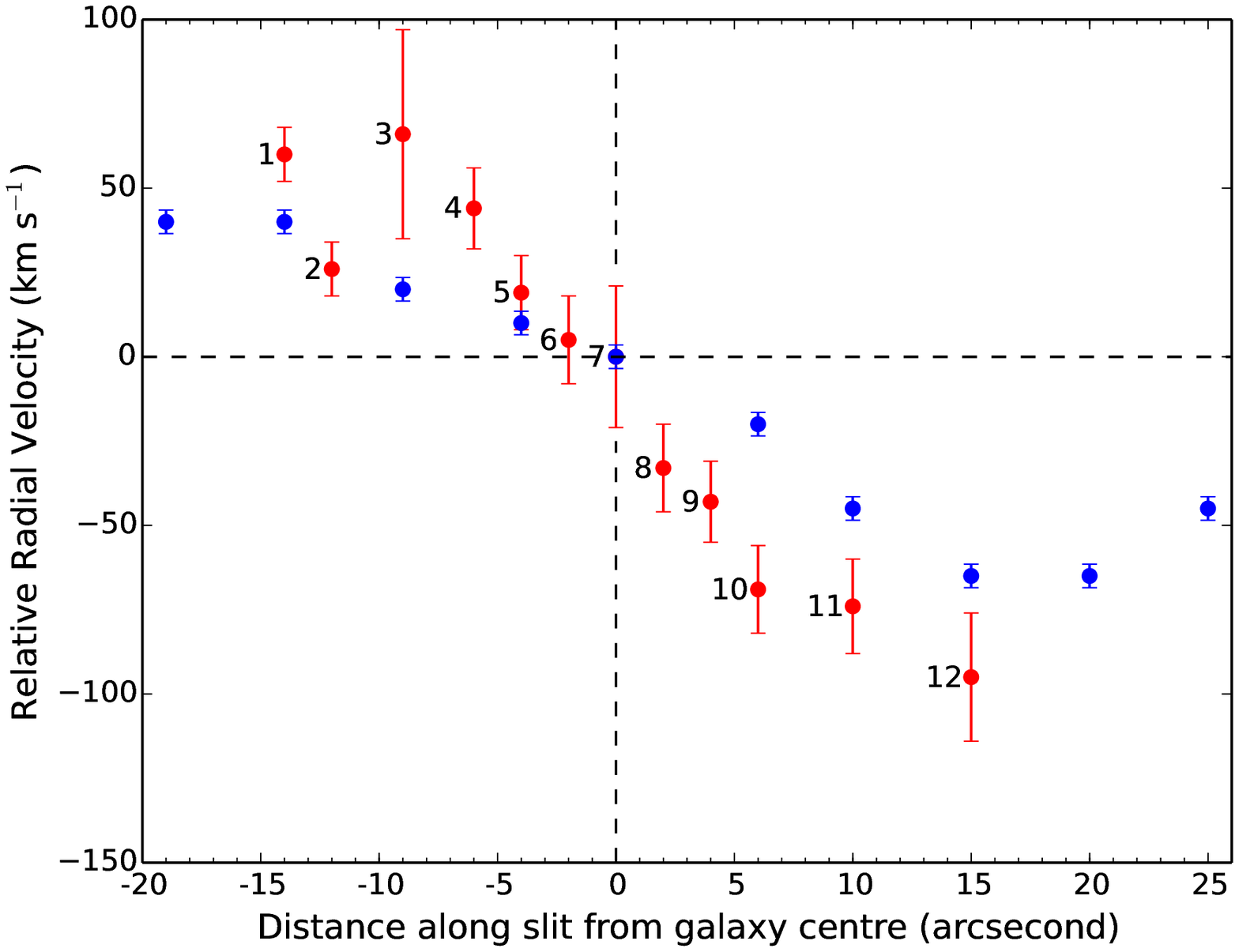}\\
{\mbox{\bf a}} &
{\mbox{\bf b}}
\end{array}$
\end{center}
\caption{\textbf{a}: Slit position and optical spectrum of the galaxy AM 0108-462. The X axis of the 2D spectrum corresponds to increasing wavelength from left to right. The position of H$\alpha$ is indicated with an arrow. \textbf{b}: Relative radial velocity of different regions of the galaxy from the gas emitting in H$\alpha$ (red data points, measured along the slit) and H{\sc i} (blue data points) wavelengths.}\label{figSpecrv}
\end{figure*}

\section{Conclusions}\label{secConclusion}

\par We observed the field of ESO 243-49 using ATCA searching for the H{\sc i} line emission in and around this galaxy. Looking for the signatures of a recent gas-rich merger event associated with the location of HLX-1 we aimed to test the scenario that HLX-1 is the stripped remnant of a dwarf galaxy that has recently undergone an interaction with ESO 243-49. We detected no H{\sc i} line emission with a 5$\sigma$ upper limit of a few 10$^8$ $M_{\odot}$. ESO 243-49 being close to the central dominant galaxy of its galaxy cluster appears to be significantly poorer in H{\sc i} gas content than the field galaxies with similar properties. The derived upper limit on the H{\sc i} line emission is consistent with an interpretation of HLX-1 being a remnant of a dwarf galaxy that has a moderate to poor H{\sc i} gas reservoir. We have ruled out the presence of a gas-rich dwarf. To investigate further more sensitive observations are needed.

\par We detected $\sim$5$\times$10$^8$ $M_{\odot}$ of H{\sc i} gas in a peculiar spiral galaxy AM 0108-462 in the field of and belonging to the same galaxy cluster as ESO 243-49. The measured H{\sc i} mass is much lower than that expected in the field for spiral galaxies with similar optical and near-infrared properties. However, AM 0108-462's H{\sc i} gas content is similar to that of the spiral galaxies at the centre of the Virgo galaxy cluster, where the dense cluster environment depletes H{\sc i} reservoirs of its member galaxies. The H{\sc i} detection in AM 0108-462 is an excellent probe of the cluster environment around ESO 243-49, providing evidence for the H{\sc i} gas reservoir depletion, therefore explaining the non-detection in ESO 243-49. The H{\sc i} line emission intensity and velocity fields revealed asymmetries including an H{\sc i} clump well outside the optical extent of AM 0108-462. We analysed archival \textit{XMM-Newton} data of an X-ray source at the centre of AM 0108-462 finding it consistent with an AGN interpretation. The \textit{Swift} UVOT data showed two ultraviolet knots of emission located along the major axis of this galaxy, and another bright knot at the location on the H{\sc i} clump pointing towards an interpretation of AM 0108-462 as an XUV disc galaxy. We further analysed optical, near-infrared and spectroscopic observations of AM 0108-462 obtained with the Magellan 6.5 m telescopes. The optical and near-infrared data showed further asymmetries in the morphology of AM 0108-462, and the spectroscopic data indicated variations in the radial velocities, suggesting that the galaxy is likely experiencing a merger event. In the merger scenario the lack of evidence in the optical spectrum of lines typical of AGN and the X-ray emission from the galaxy centre can be explained by a low-luminosity AGN that either has only recently turned on, or has a small sphere of influence (and thus a small SMBH mass).

\section*{Acknowledgments}

\par The Australia Telescope Compact Array is part of the Australia Telescope National Facility which is funded by the Commonwealth of Australia for operation as a National Facility managed by CSIRO. Parts of this research were conducted by the Australian Research Council Centre of Excellence for All-sky Astrophysics (CAASTRO), through project number CE110001020. This work is also based on observations obtained with \textit{XMM-Newton} (an ESA science mission with instruments and contributions directly funded by ESA Member States and the USA (NASA)), the 6.5 meter Magellan Telescopes located at Las Campanas Observatory, Chile, and made use of data from the \textit{Swift} Archive. S. A. Farrell was the recipient of an Australian Research Council Postdoctoral Fellowship for part of this work, funded by grant DP110102889. R. L. C. Starling is the recipient of a Royal Society Dorothy Hodgkin Fellowship.

\label{lastpage}

\bibliographystyle{mn2e}
\bibliography{bibliography}

\end{document}